\pdfoutput=1
\documentclass[%
 aip,
 jap,%
 amsmath,amssymb,
 reprint,%
]{revtex4-1}

\usepackage{graphicx}
\usepackage{dcolumn}
\usepackage{bm}
\usepackage[english]{babel}
\usepackage{placeins}
\usepackage[T1]{fontenc}
\usepackage{lmodern}
\usepackage{natbib}
\usepackage{siunitx}
\usepackage{color}
\newcommand{\etal}[1]{#1 \textit{et al.}} 

\newcommand{\comment}[1]{} 

\begin{document}

\preprint{AIP/123-QED}

\title[High spatial frequency laser induced periodic surface structure formation in germanium under strong mid-IR fields]{High spatial frequency laser induced periodic surface structure formation in germanium by mid-IR femtosecond pulses}

\author{Drake. R. Austin}
\email{austin.280@osu.edu}
\affiliation{The Ohio State University, 191 West Woodruff Ave, Columbus, OH 43210, USA}

\author{Kyle R.P. Kafka}%
\affiliation{The Ohio State University, 191 West Woodruff Ave, Columbus, OH 43210, USA}

\author{Yu Hang Lai}
\affiliation{The Ohio State University, 191 West Woodruff Ave, Columbus, OH 43210, USA}

\author{Zhou Wang}
\affiliation{The Ohio State University, 191 West Woodruff Ave, Columbus, OH 43210, USA}

\author{Kaikai Zhang}
\affiliation{The Ohio State University, 191 West Woodruff Ave, Columbus, OH 43210, USA}

\author{Hui Li}
\affiliation{The Ohio State University, 1971 Neil Ave, Columbus, OH 43210, USA}

\author{Cosmin I. Blaga}
\affiliation{The Ohio State University, 191 West Woodruff Ave, Columbus, OH 43210, USA}

\author{Allen Y. Yi}
\affiliation{The Ohio State University, 1971 Neil Ave, Columbus, OH 43210, USA}

\author{Louis F. DiMauro}
\affiliation{The Ohio State University, 191 West Woodruff Ave, Columbus, OH 43210, USA}

\author{Enam A. Chowdhury}
\affiliation{The Ohio State University, 191 West Woodruff Ave, Columbus, OH 43210, USA}

\date{\today}

\begin{abstract}
Formation of high spatial frequency laser induced periodic surface structures (HSFL) in germanium by femtosecond mid-IR pulses with wavelengths between $\lambda=$ \SIlist{2;3.6}{\micro \meter} was studied with varying angle of incidence and polarization. The period of these structures varied from $\lambda/3$--$\lambda/8$. A modified surface-scattering model including Drude excitation and the optical Kerr effect explains spatial period scaling of HSFL across the mid-IR wavelengths. Transmission electron microscopy (TEM) shows the presence of a 30 nm amorphous layer above the structure of crystalline germanium. Various mechanisms including two photon absorption and defect-induced amorphization are discussed as probable causes for the formation of this layer.
\end{abstract}


\maketitle

\section{Introduction}
Laser induced periodic surface structures (LIPSS) have been studied for decades \cite{Sipe1982} as a topic on surface science and engineering. This single step technique can produce highly ordered nano-scale features on virtually any surface from metals and semiconductors to insulators, opening the door to wide ranging applications \citep{Vorobyev2013}. Among semiconductors, such femtosecond laser processing may have applications in optoelectronics \cite{Carey2005}, solar cells \cite{Liu2014}, etc. Germanium has been gaining attention steadily due to its much wider transparency band from $\lambda =$ \SIrange{2}{17}{\micro \meter}, exceptionally high hole mobility \cite{Goley2014}, and high third order non-linearity (greater than that of silicon by almost an order of magnitude). Nano- \cite{Tang2008} and micro-structure formation in Ge \cite{Zhang2013} is of particular interest at longer wavelengths for waveguides \cite{Soref2010}, molecular sensors \cite{Tittel2003}, and integrated photonic \cite{Jalali2010,Ishikawa2010} and plasmonic devices \cite{Boltasseva2011}. Although there is a tremendous interest in mid- and far-IR light-matter interaction at present, femtosecond LIPSS work beyond \SI{2}{\micro \meter} wavelengths is almost non-existent.

We report here a systematic study of the formation of HSFL in Ge using sub-bandgap intense photon fields at wavelengths between \SIlist{2;3.6}{\micro \meter} at \SIlist{0;45;76}{\degree} angles of incidence. Similar types of HSFL formation have been reported in high bandgap materials (diamond) with laser pulses having photon energies far below the bandgap \cite{Wu2003}, where an unmodified Sipe model was used to explain the HSFL period and orientation. In this model, the incident laser light interferes with a surface-scattered wave produced during irradiation of a rough surface, leading to spatially periodic energy absorption on the surface \cite{Sipe1983}. Previous attempts by others to produce HSFL on Ge either did not succeed \cite{Borowiec2003} or required chemical etching of the surface to reveal highly disordered `HSFL' formations perpendicular to the near-IR laser polarization \cite{Harzic2013}, where an induced $\chi^{(2)}$ coupling was suggested as a generation mechanism \cite{Borowiec2003}. Low spatial frequency LIPSS (LSFL, with period $\Lambda \geq \lambda/2$) were studied previously on Ge \cite{Austin2015}, but were formed at higher fluences by the excitation of surface plasmon polaritons (SPPs) and their subsequent interference with the incident laser light \cite{Huang2009}. In this article, we present a distinct mid-IR HSFL formation mechanism based on a Sipe-Drude-Kerr (SDK) surface scattering model that takes into account electron excitation and the optical Kerr effect, resulting in excellent agreement with experimental observations. 

\section{Experimental Methods}
The experimental setup is similar to that described in detail in \etal{Austin} \cite{Austin2015}. Two OPA systems were used in this experiment to generate the range of wavelengths: 1) A Topas-C (Light Conversion) pumped by a homebuilt Ti:Sapphire system to generate \SIrange{2}{2.4}{\micro \meter} wavelength, 100 fs pulses and 2) a custom OPA to generate \SIrange{3}{3.6}{\micro \meter} wavelength, 90 fs pulses, which was also used in LSFL generation studies \cite{Austin2015}. The focal spot at each wavelength was carefully characterized using an imaging system with a mid-IR camera (Dataray, WincamD). The studied damage spots were formed with peak fluences from \SIrange{0.35}{0.38}{\joule \per \centi \meter^2}, high enough to induce the formation of HSFL, but not LSFL. The \SI{1}{\centi \meter^2} single crystal, $\langle 100 \rangle$ n-type undoped Ge samples  with resistivity $\sim 30 \: \mathrm{\Omega \cdot cm}$ were obtained from MTI Corporation. Post analysis of the damage spots was performed using scanning electron microscopy (SEM) (FEI, Helios Nanolab 600 Dual Beam), transmission electron microscopy (TEM) (FEI/Philips, CM-200T), atomic force microscopy (AFM) (Flex-Axiom, Nanosurf), and an interferometric depth profiler (IDP) (Veeco, Wyko NT9100).

\section{Results}
\subsection{Measured HSFL Periods}

\begin{figure}[t]
		\centerline{\includegraphics[width=0.48\textwidth]{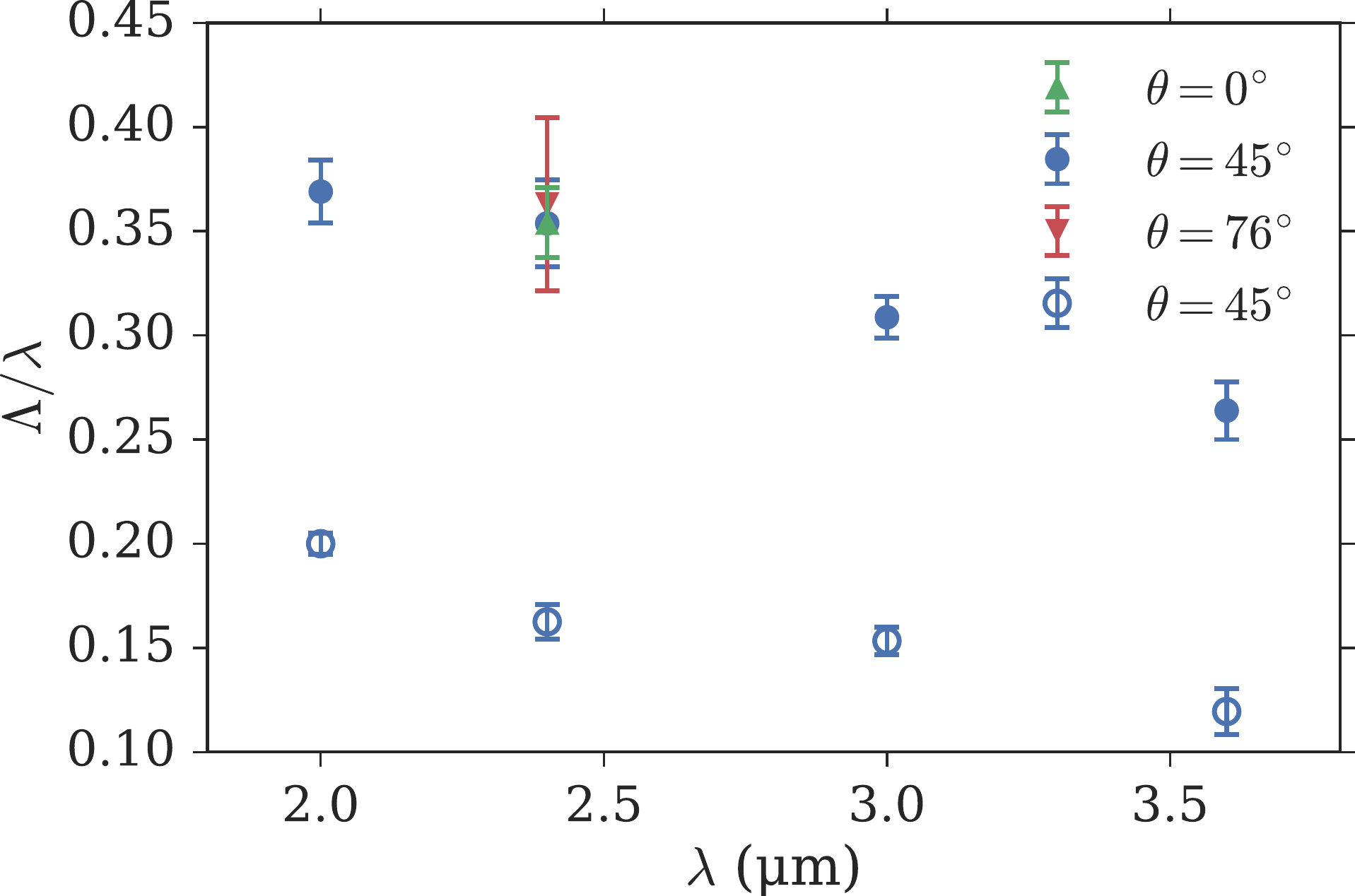}}
	\caption{Normalized HSFL period ($\Lambda/\lambda$) vs. wavelength ($\lambda$) with 100 pulses. Sipe's model of HSFL formation would predict a constant $\Lambda/\lambda$ across wavelengths; the deviations here can be explained by changes in refractive index after laser-excitation. Also plotted are the smallest observed periods of the peripheral HSFL (unfilled circles).}
	\label{fig:Period Plot}
\end{figure}

\begin{figure}
		\centerline{\includegraphics[width=0.48\textwidth]{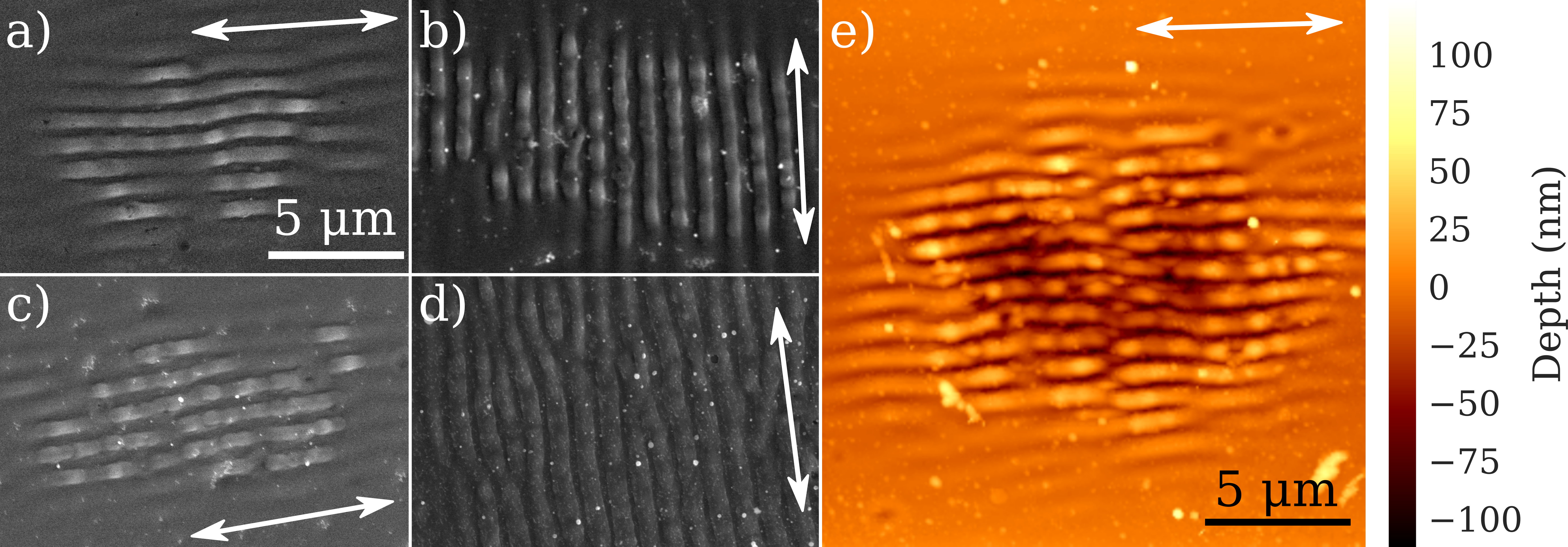}}
	\caption{(a-d) Example SEM images of central HSFL comparing \textit{p}- and \textit{s}-polarization HSFL at $\lambda =$ \SI{3.6}{\micro \meter} (a,b, respectively) and $\lambda =$ \SI{3.0}{\micro \meter} (c,d, respectively). All damage spots were formed using 100 pulses at $\theta=$ \SI{45}{\degree}. The orientation of the HSFL is found to remain parallel to the polarization (arrows). No significant difference in period is observed. (e) AFM image of central HSFL on Ge showing the surface morphology (same laser conditions as in (a)).}
	\label{fig:SEM and AFM}
\end{figure}

The HSFL period versus wavelength was recorded at $\lambda=$ \SIlist{2;2.4;3;3.6}{\micro \meter} wavelengths under $\theta =$ \SI{45}{\degree} illumination, whereas the angular dependence was obtained at \SIlist{0;45;76}{\degree} at $\lambda=$ \SI{2.4}{\micro \meter}. HSFL were observed to form both in the central region of the damage spot as well as in the periphery, though never in the intermediate region. HSFL were oriented parallel to the laser polarization, ruling out second harmonic generation as a probable cause \cite{Harzic2013,Dufft2009}. The central HSFL periods were determined by taking the 2D Fourier transform of SEM and IDP images and identifying peaks in the spectra whereas the periphery HSFL periods were determined by taking multiple lineouts. The primary experimental results are presented in Fig. \ref{fig:Period Plot}, showing the measured period (normalized to the laser wavelength) as a function of wavelength for 100 pulses, revealing an approximately linear dependence and no noticeable dependence on $\theta$. The central and peripheral HSFL are represented by filled and unfilled markers, respectively. The periods of peripheral HSFL decreased with increasing distance from the center, possibly due to the decrease in local fluence. In Fig. \ref{fig:Period Plot}, the shortest periods of the peripheral HSFL are presented for simplicity; they were found to be approximately half the period of the central HSFL. For the remainder of this paper, the focus will be on the central HSFL as the origin of the peripheral HSFL is not yet clear, though it could be related to the presence of a native oxide layer which has been observed to result in an outer damage ring surrounding a central damage spot on Si \cite{McDonald2005}. Example HSFL images are also shown in Fig. \ref{fig:SEM and AFM}(a-d), comparing \textit{s}- and \textit{p}-polarizations at \SIlist{3;3.6}{\micro \meter} wavelengths. In all cases, the orientation of the HSFL remains parallel to the polarization. Despite the strong polarization dependence on $\Lambda$ in the case of LSFL \cite{Austin2015}, no significant variation with polarization is observed here with central HSFL.

\subsection{Surface Morphology}
\begin{figure*}
		\centerline{\includegraphics[width=1.0\textwidth]{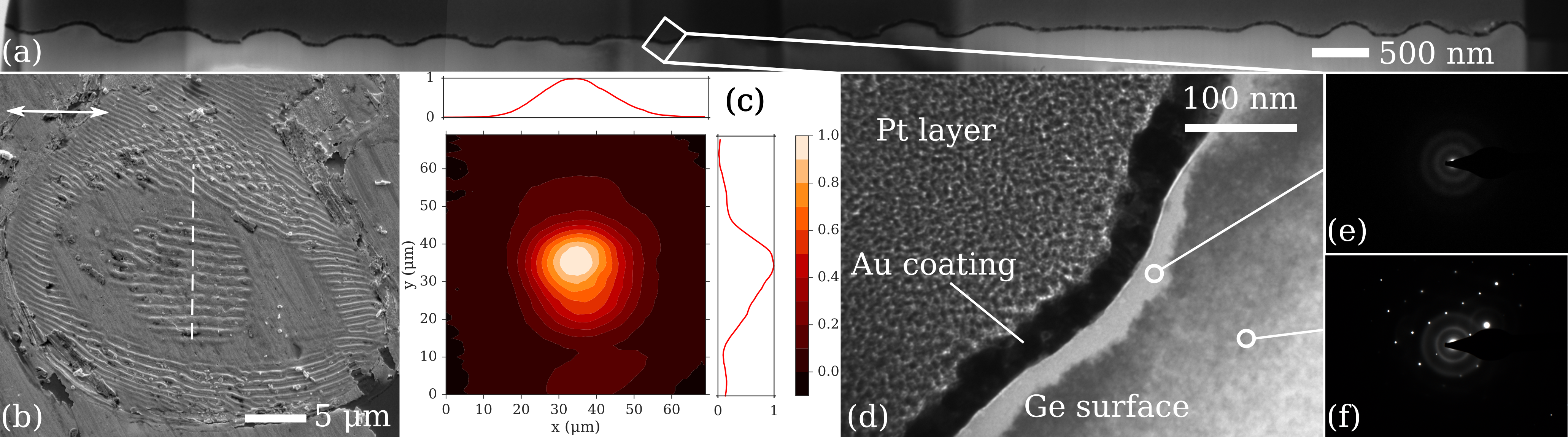}}
	\caption{(a) TEM cross section showing the structure of central HSFL on Ge ($\lambda =$ \SI{2.4}{\micro \meter}, $\theta=$ \SI{0}{\degree}, 100 pulses, \SI{0.36}{\joule \per \centi \meter^2}). (b) SEM image of the damage spot chosen for cross-section TEM imaging (arrow indicates polarization direction). (c) Focal spot profile of the \SI{2.4}{\micro \meter} wavelength beam, ruling out large intensity variations in the focal spot as a probable cause for the absence of HSFL in the intermediate region. (d) High magnification TEM image of the Ge surface. Beneath the gold coating (black) is an oxide layer $\sim$ 3 nm thick followed by an amorphous top layer of Ge capping bulk crystalline Ge. This amorphous layer is present throughout the cross-section. (e) Diffraction pattern of electrons transmitting through the brighter amorphous layer, showing no crystallinity. (f) Same as (e) but through the crystalline bulk, confirming the crystallinity. [See supplementary material for more details in (a) \cite{SuppMat}.]}
	\label{fig:HSFL TEM}
\end{figure*}

A cross-sectional specimen of the laser spot shown in Fig. \ref{fig:HSFL TEM}(b) was prepared using a focused ion beam \cite{Langford2001} (FEI Nova NanoLab 600 Dual Beam). An \textit{in-situ} transfer was performed by attaching the specimen to a micromanipulator through platinum deposition. A final cut was then made, separating the specimen from the sample. After attaching the specimen to the TEM grid through additional platinum deposition, a final thinning was performed until reasonable electron transparency was achieved. Additionally, because the ion-assisted platinum layer deposition can result in the amorphization of the surface down to approximately \SI{30}{\nano \meter} \cite{Langford2001}, a thin ($\sim 40$ nm) layer of gold was first deposited onto the sample using \SI{1}{\kilo \volt} DC sputtering, protecting the surface of the Ge from the deposition process. Otherwise, any modifications to the surface observed under TEM, especially the formation of an amorphous layer, would not readily be attributed to laser exposure. Fig. \ref{fig:HSFL TEM} shows select TEM images of Ge HSFL formed at the center of the damage spot ($\lambda=$ \SI{2.4}{\micro \meter}, $\theta=$ \SI{0}{\degree}, 100 pulses). The observed period at this wavelength was $850 \pm 50$ nm with a peak-to-trough height of $115 \pm 25$ nm. AFM images (Fig. \ref{fig:SEM and AFM}(e)) show similar surface morphology at \SI{3.6}{\micro \meter} with shallower ripples further away from the center of the damage spot. Capping the surface of the crystalline Ge is an amorphous layer $\sim 30$ nm thick, similar to the depths reported in GaP,  InP, Si, and SiC \cite{Hsu2008, Couillard2007, Schade2009, Reif2010, Tomita2010}. The formation of these layers has been attributed to the melting and subsequent rapid resolidification of the surface into a highly disordered structure after laser exposure \cite{Couillard2007}.

\section{Discussion}
\subsection{Description of Theoretical Model}
\begin{figure}
		\centerline{\includegraphics[width=0.45\textwidth]{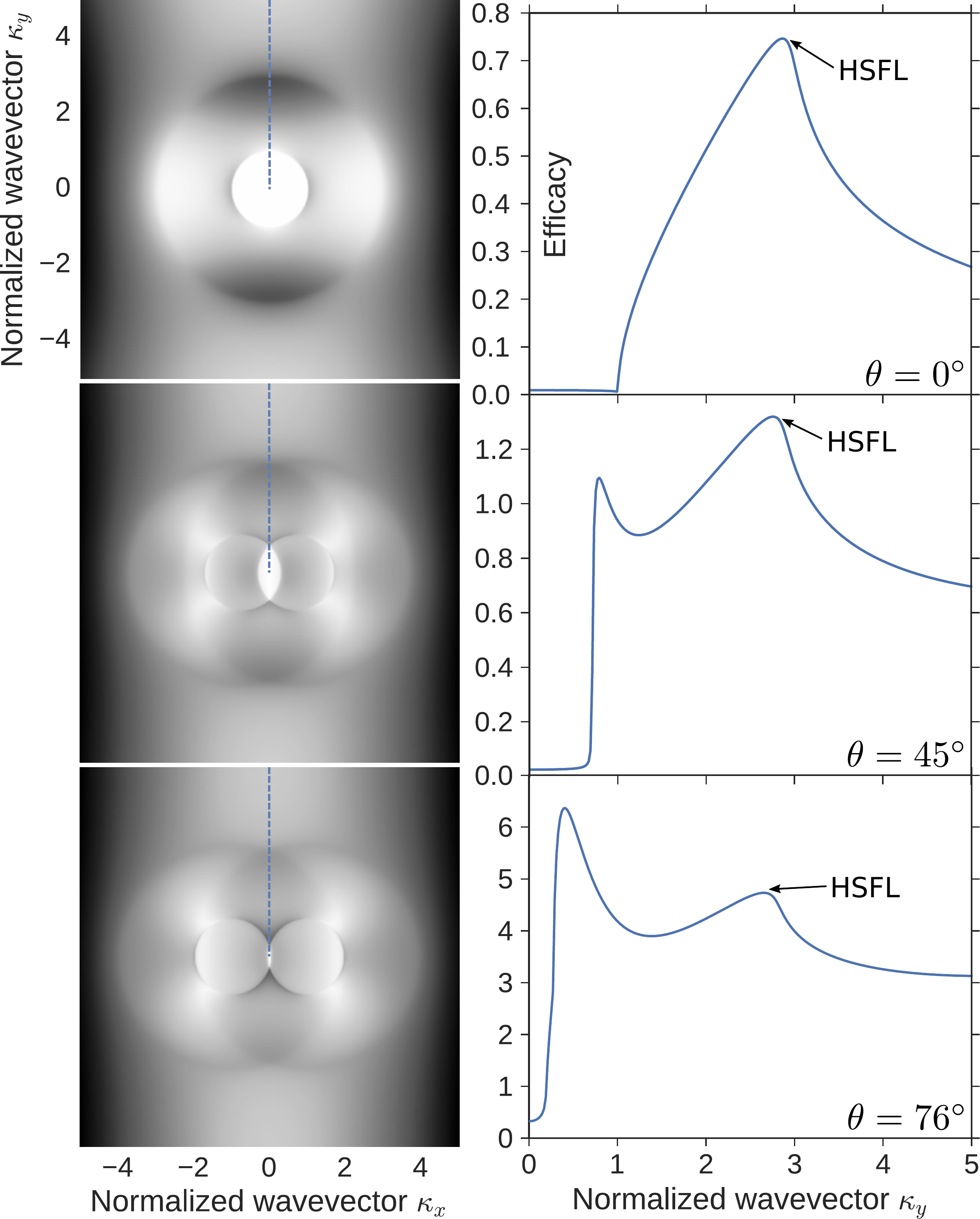}}
	\caption{Left: Plots of the efficacy factor $\eta (\mathbf{k})$ at $\lambda =$ \SI{2.4}{\micro \meter}, $\theta$ = \SIlist{0;45;76}{\degree} with $n_e = 2.41 \times 10^{20}$ cm$^{-3}$. Right: Vertical lineouts through the center. Peaks in this spectrum correspond to surface wave vectors that exhibit strong absorption and can therefore lead to LIPSS formation with orientation parallel to the laser polarization. A strong peak near the observed HSFL period is present, showing negligible variation with $\theta$ as observed experimentally. The peak is, however, observed to weaken at \SI{76}{\degree}, explaining the greater uncertainty in the period at that angle.}
	\label{fig:Efficacy}
\end{figure}

HSFL formation was modeled using the theory of \etal{Sipe} \cite{Sipe1983} in which the interference of the incident laser pulse with a surface scattered wave results in the inhomogeneous absorption of energy. The efficacy at which this inhomogeneity occurs is described by the function $\eta (\mathbf{k})$ where $\mathbf{k}$ is the surface wave vector. This function exhibits peaks at particular wave vectors (Fig. \ref{fig:Efficacy}); an initially rough surface with Fourier components at these peaks will have them reinforced with each laser pulse. \etal{Bonse} \cite{Bonse2005} derived a series of equations that can be used to calculate this efficacy factor given the laser wavelength, polarization, angle of incidence, material permittivity, and the surface shape and filling factors that describe the surface roughness. Here, the shape and filling factors were chosen to be 0.4 and 0.7, respectively, as these were the values for Ge that best matched the original data reported by \etal{Sipe} \cite{Young1983}.

The complex material permittivity used was a combination of the non-excited value for Ge at the specified wavelength ($\epsilon_c \approx 16$ for mid-IR wavelengths)
\begin{equation}\label{Efficacy}
\epsilon = \epsilon_c + \epsilon_{Drude} + \epsilon_{Kerr} 
\end{equation}
together with modifications due to the Kerr effect as well as laser excitation according to the Drude model \cite{Dufft2009}
\begin{equation}
\epsilon_{Drude} = - \frac{\omega_p^2}{\omega (\omega + i \Gamma)}, \: \epsilon_{Kerr} = \frac{3 \chi^{(3)}I}{2 n_0 c \epsilon_0} = 2 n_0 n_2 I.
\end{equation}
Here, $\omega_p = \sqrt{e^2 n_e / m^* \epsilon_0}$ is the plasma frequency, $n_e$ is the conduction band electron density, $m^*$ is the optical effective mass, and $\Gamma$ is the electron collision frequency. The electron density was used as a free parameter corresponding to varying amounts of laser-excitation while values used for the optical effective mass and electron collision frequency were $m^* = 0.081 m_e$ and $1/\Gamma = 46$ fs, respectively. The latter value was taken from \etal{Austin} \cite{Austin2015} and corrected for the Kerr effect, which was not considered in that paper. It should be noted that this expression for $\epsilon_{Kerr}$ does not assume $\epsilon_{Kerr} << n_0^2$ as was done by \etal{Dufft} \cite{Dufft2009}. The third-order susceptibility $\chi^{(3)}$ of Ge at each wavelength (and, consequently, the Kerr coefficient $n_2$) was taken from the theoretical dispersive curve presented by \etal{Hon} \cite{Hon2011} that provided the best fit to the experimental data at mid-IR wavelengths. The Kerr effect cannot be neglected here for Ge as the values of $n_2$ are particularly high (0.38, 2.51, 2.07, and 1.46 $\times 10^{-13} \; \mathrm{cm^2/W}$ for \SIlist{2.0;2.4;3.0;3.6}{\micro \meter} wavelengths, respectively), significantly influencing HSFL formation.

\subsection{Angular and Polarization Dependence}
An example plot of the efficacy factor for \SI{2.4}{\micro \meter} wavelength, \textit{p}-polarized light at $\theta$ = \SIlist{0;45;76}{\degree} is shown in Fig. \ref{fig:Efficacy} with an electron density of $2.41 \times 10^{20}$ cm$^{-3}$. The surface wave vector has been normalized to the laser wavelength. Multiple peaks are apparent, some with large periods corresponding to LSFL (not observed) and a peak at the observed HSFL period with orientation parallel to the laser polarization (see vertical lineouts in Fig. \ref{fig:Efficacy}). The location of this HSFL peak shows negligible variation with $\theta$ and no variation when the polarization is changed, as observed experimentally (Fig. \ref{fig:Period Plot}). This is inconsistent with the model of LIPSS formation in which SPPs are excited on the metallized surface and interfere with the incident laser light, which has been successfully used to explain observed properties of LSFL \citep{Austin2015, Huang2009}. For \textit{s-} and \textit{p-}polarized light, the respective LIPSS periods predicted by this SPP model are given by
\begin{equation}
\Lambda_s = \frac{\lambda}{\sqrt{(\lambda/\lambda_s)^2 - \sin^2 \theta}}, \: \Lambda_p = \frac{\lambda}{\lambda/\lambda_s - \sin \theta}
\end{equation}
where $\lambda_s$ is the SPP wavelength. However, the fact that typically $\lambda/\lambda_s \approx 1$ leads to a strong dependence on polarization and angle of incidence. It is therefore unlikely that the HSFL observed here are a result of SPP excitation. Additionally, the previously mentioned LSFL were only observed to form at higher fluences ($\gtrsim$ \SI{0.4}{\joule \per \centi \meter^2}) while HSFL were observed to form at lower fluences ($\lesssim$ \SI{0.4}{\joule \per \centi \meter^2}) and in the peripheries of damage spots. This is consistent with the requirement that $\mathrm{Re}[\epsilon] < -1$ in order for SPPs to form, a condition that is not satisfied until fluences high enough to cause sufficient ionization are achieved. Below this fluence, the surface remains non-metallic and the usual surface-scattered-wave-induced LIPSS dominate. 

A similar analysis can be performed with analytic expressions used to model LIPSS formation \cite{Csete2001} as described by \etal{Sipe} \cite{Sipe1983}:
\begin{equation}
\Lambda_s = \frac{\lambda}{n - \sin \theta}, \: \Lambda_p = \frac{\lambda}{\sqrt{n^2 - \sin^2 \theta}}.
\end{equation}
For the case of Ge at mid-IR wavelengths, $n \approx 4 >> \sin \theta$ and $\Lambda_s \approx \Lambda_p \approx \lambda/n$, yielding little dependence on polarization or angle of incidence, as observed. Additionally, a linear dependence on wavelength would be predicted for constant $n$, which was not observed in Fig. \ref{fig:Period Plot} due to deviations from linearity from changes in $n$ after laser excitation.

\subsection{Electron Density Estimates}

\begin{figure}
		\centerline{\includegraphics[width=0.48\textwidth]{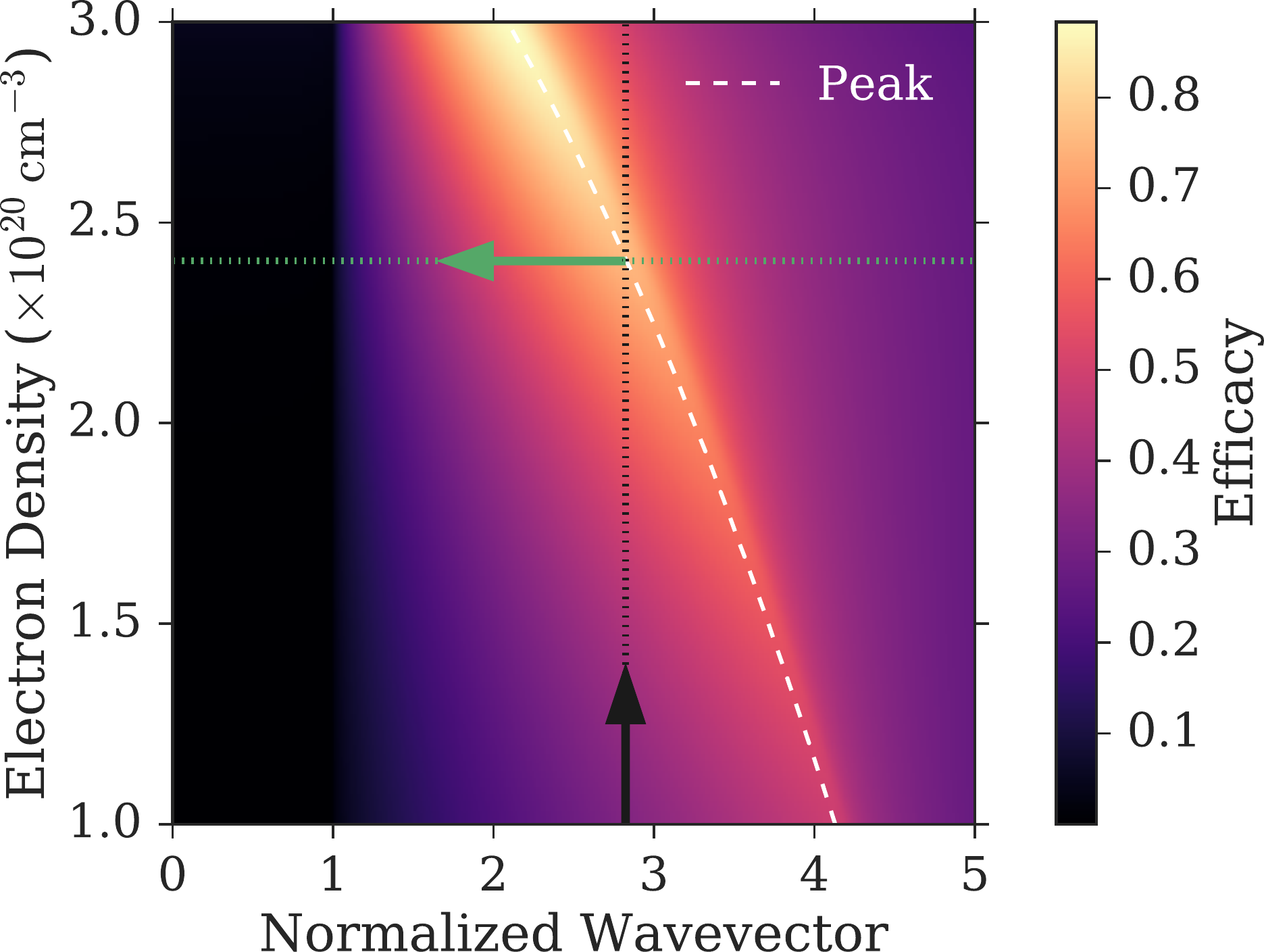}}
	\caption{Variation of efficacy factor with surface wavevector and electron density for $\lambda =$ \SI{2.4}{\micro \meter}, $\theta=$ \SI{0}{\degree}, 100 pulses. The white dashed curve traces out the peak of the efficacy for each electron density. The black dotted line corresponds to the observed HSFL period while the green dotted line corresponds to the electron density required for the efficacy peak to match this period.}
	\label{fig:Efficacy ne dependence}
\end{figure}

\begin{figure}
		\centerline{\includegraphics[width=0.48\textwidth]{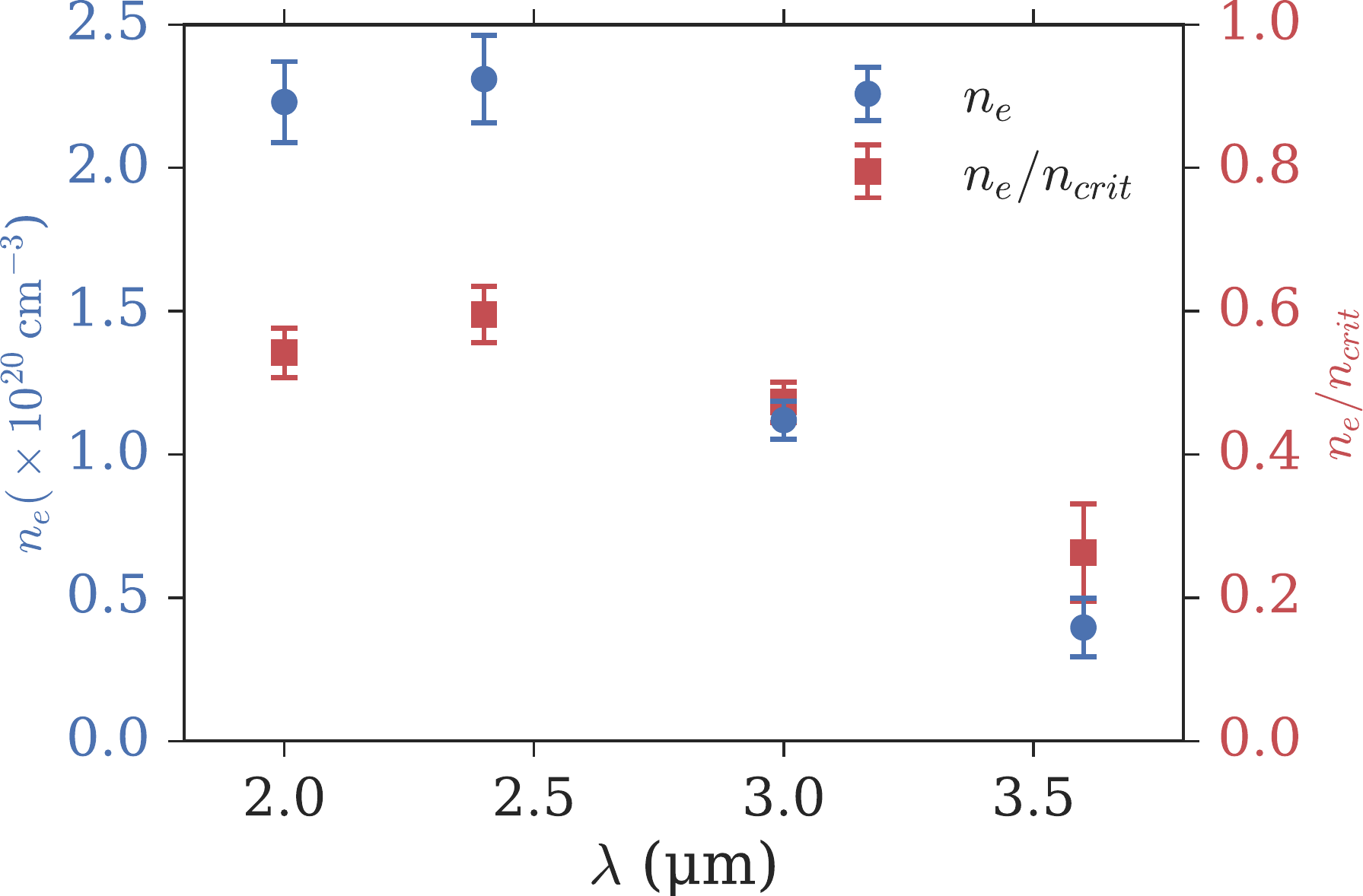}}
	\caption{Fitted electron density as a function of wavelength (100 pulses, $\theta =$ \SI{45}{\degree}). The right axis has been scaled to a critical density defined as the electron density at which the real part of Eq. \eqref{Efficacy} equates to zero.}
	\label{fig:ne vs wavelength}
\end{figure}

Efficacy plots similar to those in Fig. \ref{fig:Efficacy} were made for each set of laser conditions, using the electron density as a fitting parameter. Fig. \ref{fig:Efficacy ne dependence} summarizes this process by showing the variation of the efficacy factor with wavevector and electron density for the laser conditions that produced the central HSFL in Fig. \ref{fig:HSFL TEM}. The peak of the efficacy factor is denoted by a white dashed line and is observed to shift to smaller wavevectors as the electron density increases. The black dotted line denotes the wavevector at which HSFL were observed to form; by tracing this line to the efficacy peak, a predicted value for the electron density can be extracted (green dotted line). This process was repeated at each wavelength for $\theta=$ \SI{45}{\degree} with \textit{p}-polarization; the results are plotted in Fig. \ref{fig:ne vs wavelength}. The right axis is normalized to a surface critical density $n_{crit}$ defined as the electron density at which the real part of Eq. \eqref{Efficacy} equates to zero, when the solid surface becomes metallic. In all cases, the observed HSFL wavevector could be matched using reasonable values of electron density (near critical density). In general, it appears that these central HSFL tend to form at $n_e \approx n_{crit}/2$. This is in contrast to the electron density found in \etal{Austin} \cite{Austin2015} based on LSFL analysis at \SI{3.0}{\micro \meter}, which, after correcting for the Kerr effect, is $2.82 \times 10^{20} \mathrm{cm^{-3}}$ ($1.19 n_{crit}$). This higher electron density for LSFL is to be expected as a higher fluence was used (0.43 vs \SI{0.36}{\joule \per \centi \meter^2}) and a metallic surface is required for SPPs to be excited. The coupling of energy to the surface is stronger at these higher fluences, resulting in a $\sim$ \SI{1}{\micro \meter} ripple depth as opposed to the $\sim$ \SI{100}{\nano \meter} ripple depths observed here with HSFL.

Because the HSFL periods were measured by taking the FFT of the central region, it's worth noting that these electron densities should be taken as averages over the region in which central HSFL were observed to form. While the peak fluence was used in the calculation of the electron densities, using the average fluence of the central HSFL region changes the predicted value by $\sim 2 \%$, so the former was used to simplify the calculations. In principle, the decrease in local fluence from the center outward results in a decrease in electron density and, therefore, a variation in HSFL period. However, because the Kerr effect opposes the effect of Drude excitation, the HSFL period is less sensitive to changes in fluence than it would be otherwise, particularly for Ge with its large $\chi^{(3)}$. In addition to this, the central HSFL regions are relatively small with only modest variations in local fluence. For example, the decrease in fluence in Fig. \ref{fig:HSFL TEM}(a) from the center of the damage spot to the outermost edge of the central HSFL region is less than 10\%. As a result, the variations in HSFL period within the central region are difficult to distinguish from the stochastic fluctuations inherent to LIPSS formation.

\subsection{Amorphous Layer}
For the parameters of Fig. \ref{fig:HSFL TEM}, the electron density is found to be $2.41 \times 10^{20} \: \mathrm{cm^{-3}}$, which can be used to construct a simple model to analyze the observed laser-induced amorphization. This was performed by modeling the laser pulse as Gaussian in time (accounting for the change in the Kerr effect with intensity throughout the pulse) and by assuming the decay of the electric field strength is due to absorption. However, under the specified laser conditions, the energy absorbed never exceeds 0.2 eV/atom, well below the energy needed for melting of 0.6 eV/atom (this includes the enthalpy of fusion of 0.38 eV/atom for Ge). The thermal accumulation due to multiple pulses was also found to be negligible due to the large pulse separation (1 ms) \citep{Harzic2013_2}. The formation of this amorphous layer therefore cannot be attributed to Fresnel absorption. Instead, it is necessary to consider two-photon absorption (TPA) according to the expression $\mathrm{d}I(z) / \mathrm{dz} = \beta I^2$ where $\beta$ is the TPA coefficient and $I$ is the intensity. In order to obtain an accurate estimate of the energy absorbed due to this effect, a value of $\beta$ at a \SI{2.4}{\micro \meter} wavelength in the femtosecond regime with high intensities ($\sim \mathrm{TW/cm^2}$) is needed, which is currently lacking. A rough estimate can be made by using the values of $\beta=$ \SI{80}{\centi \meter \per \giga \watt} reported by \etal{Rauscher} \cite{Rauscher1997} for \SI{2.9}{\micro \meter} light in the picosecond regime with intensities up to \SI{2}{\giga \watt \per \centi \meter^2}. Scaling to a \SI{2.4}{\micro \meter} wavelength based on photon energy yields $\beta=$ \SI{66}{\centi \meter \per \giga \watt}. Using this value to determine the energy absorption as a function of depth suggests the formation of a melted layer extending down to \SI{135}{\nano \meter}. However, values of $\beta$ have been known to decrease with pulse duration \cite{Dragonmir2002} which would reduce this estimate. 

Another possible mechanism of amorphous layer formation is the formation of defect states after exposure to multiple laser pulses. While a disordered lattice configuration would be entropically favorable, the lower internal energy of an ordered lattice more than makes up for this difference in entropy when the crystal is below the melting temperature. With the introduction of defects, however, the internal energy of the crystalline phase can be increased until the material changes to an amorphous phase in order to lower its Gibbs free energy \cite{Fecht1992}. How the depth of the amorphous layer changes as a function of the number of pulses would allow better understanding of this process.

\section{Conclusion}
In summary, the formation of central HSFL on Ge at mid-IR wavelengths is consistent with an SDK surface-scattered wave model of LIPSS formation. This is in contrast to the LSFL formation mechanism in the same wavelength regime \cite{Austin2015} where higher fluences generate a metallic surface layer, allowing for the excitation of SPPs and their subsequent interference with the incident laser light. The inclusion of Drude excitation in the SDK model allows for an estimate of the electron density after laser-excitation. These estimates were significantly influenced by the Kerr effect because of the high third-order susceptibility of Ge. With these effects taken into account, it was found that central HSFL on Ge seem to form optimally when the electron density is approximately half of the surface critical density. Finally, two possible mechanisms were introduced to qualitatively explain the formation of the amorphous layer in the HSFL region. To identify the mechanism for quantitative agreement with observations, further studies at different pulse numbers as well as a measurement of the TPA coefficient for Ge in the mid-IR and femtosecond regimes at $\mathrm{TW/cm^2}$ intensities are needed.

\FloatBarrier
\section{Acknowledgments}
This material is based upon work supported by the Air Force Office of Scientific Research (AFOSR), USA under award no. FA9550-12-1-0454, FA9550-12-1-0047, and FA9550-16-1-0069 as well as the Air Force Research Laboratory, USA award no. FA-9451-14-1-0351. The DiMauro group acknowledges support from MIR MURI award FA9550-16-1-0013. C. I. Blaga acknowledges support from AFOSR YIP, award FA9550-15-1-0203.
\newline

\FloatBarrier
\renewcommand{\section}[2]{}%
\bibliographystyle{apsrev4-1}
\bibliography{Germanium_HSFL_Paper_JAP}

\widetext
\pagebreak
\begin{center}
\textbf{\large Supplementary Material: High spatial frequency laser induced periodic surface structure formation in germanium by mid-IR femtosecond pulses}
\end{center}

\renewcommand{\thefigure}{S1}
\begin{figure*}[h]
		\centerline{\includegraphics[width=1.0\textwidth]{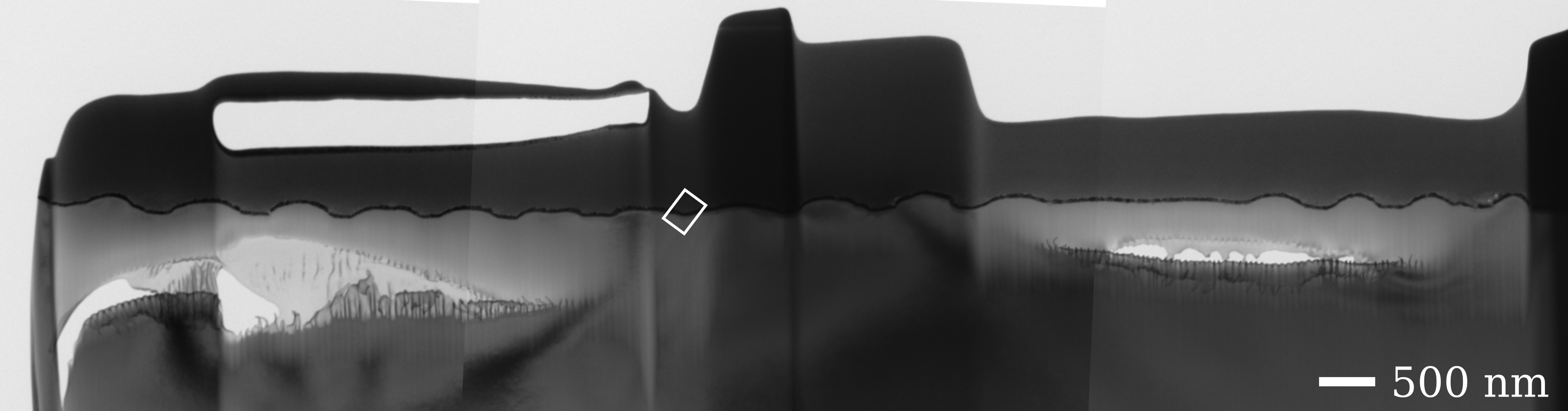}}
	\caption{Enlarged cross-section shown in Fig. \ref{fig:HSFL TEM}(a). The boxed region shows where the higher resolution image in Fig. \ref{fig:HSFL TEM}(c) was taken. The visible gaps are where the sample was over-thinned.}
	\label{fig:TEM Supplementary}
\end{figure*}

\end{document}